\documentclass[aps,prb,amsfonts,amssymb,twocolumn,amsmath,preprintnumbers,nofootinbib,floatfix,
showpacs]{revtex4-1}
\usepackage[dvips]{graphics}
\usepackage{graphicx}
\usepackage{bm}
\begin{document}

\title{Injection of nonequilibrium quasiparticles into Zeeman-split superconductors: a way to create long-range spin imbalance}
\author{I. V. Bobkova}
\affiliation{Institute of Solid State Physics, Chernogolovka,
Moscow reg., 142432 Russia}
\affiliation{Moscow Institute of Physics and Technology, Dolgoprudny, 141700 Russia}
\author{A. M. Bobkov}
\affiliation{Institute of Solid State Physics, Chernogolovka,
Moscow reg., 142432 Russia}

\date{\today}

\begin{abstract}
A theory of spin transport and spin detection in Zeeman-split superconducting films at low temperatures is developed. It is shown that an injection of spin-unpolarized quasiparticles into a Zeeman-split superconductor gives rise to a spin imbalance. The relaxation length of such a spin signal is determined by the energy relaxation length and can be extremely large as compared to the relaxation length of spin-polarized quasiparticles.
There can exist two types of signals: due to nonthermalized quasiparticle distribution and due to thermalized overheated electron distribution. They have different decay lengths and can be distinguished by their different dependencies on the applied voltage. 
The decay length of the nonthermalized signal is determined by the electron-electron scattering rate, renormalized  due to superconductivity. The decay length of the thermalized signal is determined by the length on which energy leaves the electronic subsystem and can be very large under special conditions. 
Applications of the theory to recent experimental data on spin relaxation in Zeeman-split and exchange-split superconductors are discussed. In particular, it can explain the extremely high spin relaxation lengths, experimentally observed in Zeeman-split superconductors, and their growth with the magnetic field and with the applied voltage. 
\end{abstract}
\pacs{74.78.Na, 74.25.fg, 85.75.-d}

\maketitle

\section{introduction}

The superconducting spintronics now is a very active field of research. It is based on the recent progress in realization of superconductor/ferromagnet heterostructures. One direction of the research activity has been focused on the study of proximity induced triplet superconducting correlations in equilibrium \cite{buzdin05,bergeret05,keizer06,robinson10,khaire10,anwar10,bergeret14,gomperud15,jacobsen15_1,jacobsen15_2}. The other direction is to study spin-polarized quasiparticle transport and spin accumulation in superconducting wires \cite{giazotto08,yang10,poli08,hubler12,quay13,wolf13,wakamura14,wolf14,silaev14,virtanen15}. In particular, it is very important to transmit spin signals over mesoscopic length scales. 

Usually spin signals are created by injection of spin-polarized quasiparticles into normal or superconducting wire from ferromagnetic leads. It was shown in transport experiments \cite{jedema02,hubler12,quay13} that for Al thin films in the normal state the spin relaxation length $\lambda_N$ is of the order of $400-500$ nm. It was also measured that the spin relaxation length is reduced upon Al transition into the superconducting state \cite{poli08}. 

However, it was demonstrated recently that in Zeeman-split superconducting films the spin signals can be created by injection of unpolarized electrons. Such spin signals can spread over distances of several $\mu$m \cite{hubler12,quay13,wolf13,wolf14}. In these experiments the spin relaxation length exceeds considerably the superconducting coherence length, the normal-state spin relaxation length and the charge-imbalance length. Moreover, the decay length of the spin signal grows with the applied magnetic field, while the charge-imbalance relaxation length only reduces.

The origin of such long-range spin signals has been addressed by several theoretical groups recently. It is known that in the absence of the magnetic or the exchange field (Zeeman splitting of the DOS) and at low temperatures the main mechanisms of the spin relaxation in superconductors are elastic spin flips by magnetic impurities and by spin-orbit interaction \cite{zhao95,morten04,morten05,poli08,yang10}. In order to these mechanisms of spin relaxation can work there should be a difference between distribution functions for quasiparticles with opposite spins. It has been shown \cite{bobkova15,silaev15} that for realistic parameters of the films the relaxation length provided by these mechanisms in the Zeeman-split superconducting state does not exceed the normal state relaxation length, which is of the order of $400-500$ nm in the experimentally investigated films. So, it is unlikely that the experimentally observed long-distance spin relaxation is provided by such elastic spin-flip processes.      

Instead, it was proposed \cite{bobkova15,silaev15,krishtop15} that for the observed spin signal a difference between distribution functions for quasiparticles with opposite spins is absent. But the quasiparticle current injected into the superconductor is accompanied by the energy flow that creates a nonequilibrium quasiparticle distribution in it. The role of elastic spin-flip processes is only to rapidly relax the distribution function to the spin-independent value. The observed spin signal is formed by this spin-independent nonequilibrium quasiparticle distribution weighted by the spin-split DOS.  The relaxation length of such a spin signal is the energy relaxation length. 

In the framework of this mechanism the shape of the spin signal is well reproduced \cite{bobkova15,silaev15}. However, there is no full and detailed theoretical investigation of the long-range spin signal relaxation so far. In Refs.~[\onlinecite{krishtop15}] and [\onlinecite{silaev15}] the relaxation length was considered as a phenomenological parameter, in Ref.~[\onlinecite{bobkova15}] the renormalization of the energy relaxation time due to superconductivity was not taken into account and the overheating of the electron subsystem was neglected. 

In the present paper we develop a theory of spin relaxation in Zeeman-split superconducting films at low temperatures and focus on the relaxation mechanisms of the long-range spin signal. It has been reported in the literature \cite{kopnin09,moor09} that the energy relaxation provided by the electron-electron scattering in Al at low temperatures is faster than the relaxation due to electron-phonon scattering. In this framework we show that the long-range spin signal can be naturally divided into two parts: due to nonthermalized quasiparticle distribution and due to thermalized overheated electron distribution. They can be distinguished by their different dependencies on the applied voltage. 
The decay length of the nonthermalized signal is determined by the electron-electron scattering rate, renormalized  due to superconductivity. It depends crucially on the temperature of the overheated electron subsystem and superconducting gap. The decay length of the thermalized signal is determined by the length on which the injected energy leaves the electronic subsystem. In dependence on the particular sample design it can be determined by an electron-phonon relaxation length or the geometry, that is the distance between the injector and the heat reservoir. In realistic systems this length can be very large. 

Applications of the theory to the recent experimental data \cite{hubler12,wolf13,wolf14} on spin relaxation in Zeeman-split and exchange-split superconductors are discussed. In particular, it can explain the extremely high spin relaxation lengths, experimentally observed in Zeeman-split superconductors, their growth with the magnetic field and with the applied voltage. It also reproduces the characteristic two-peak shape of the signal, measured for the exchange-split samples \cite{wolf14}. 

The paper is organized as follows.  In Sec.~\ref{model} we describe the system under consideration and discuss qualitatively the physics of the effect. In Sec.~\ref{approach} the developed theoretical approach is formulated. In Sec.~\ref{results} the results of our calculations are presented, discussed and compared to the experimental data. Our conclusions are given in Sec.~\ref{conclusions}. 

\section{model and qualitative description of the effect}

\label{model}

\begin{figure}[!tbh]
  \centerline{\includegraphics[clip=true,width=3.5in]{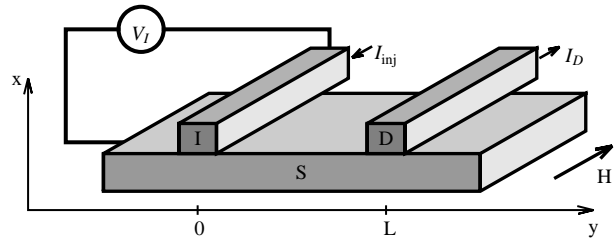}}
   \caption{Scheme of the system under consideration.}   
\label{scheme}
\end{figure}

Following the experiments \cite{hubler12,quay13,wolf13} we consider the system depicted in Fig.~\ref{scheme}. It consists of a thin superconducting film (S) overlapped by the injector (I) and detector (D) electrodes. The distance between them is $L$. Both the injector and the detector are coupled to the film by tunnel contacts. A current is injected into the superconducting film via I. This electrode can be normal or ferromagnetic. The detector electrode is ferromagnetic. In this case the spin imbalance in the superconductor can be converted into an electric current at the S/D interface. For the tunnel case, considered here,
the current $I_D$ measured by the detector can be calculated as
\begin{equation}
I_D=G_D(\mu + P_D S)
\label{ID}
\enspace ,
\end{equation}
where $G_D=G_\uparrow +G_\downarrow $ is the total conductance of the S/D interface and $P_D=(G_\uparrow -G_\downarrow)/G_D$ is its polarization. 
$\mu$ in the right hand side of Eq.~(\ref{ID}) is the shift of quasiparticle chemical potential, determined by the charge
imbalance, and the second term in the brackets is proportional to the local nonequilibrium spin accumulation $S$ in the film at the detector point. Further we are interested in large enough $L$, where the charge imbalance has already relaxed: $\mu=0$. So, we assume that the electric current at the detector point is proportional to $S$ and focus on this quantity. In the real experimental situation the charge imbalance indeed relaxes much more rapidly than the spin imbalance \cite{hubler12,quay13,wolf13}. Physically, the main source of the low-temperature charge imbalance relaxation is the orbital effect of the applied magnetic field \cite{nielsen82,silaev15}. The magnetic field is applied in plane of the film and is parallel to the ferromagnetic wires. In our study the quantization axis is chosen along the magnetic field.   

Now we discuss the qualitative physics lying behind these measurements. As it was already mentioned in the introduction, it is desirable to make the spin relaxation length as large as possible. In the absence of the magnetic or the exchange field (Zeeman splitting of the DOS) and at low temperatures the main mechanisms of the spin relaxation in superconductors are elastic spin flips by magnetic impurities and by spin-orbit interaction \cite{zhao95,morten04,morten05,poli08,yang10}. The resulting relaxation length is not large: of the order of a few hundreds nanometers for Al. It was also shown experimentally and theoretically that it can be only reduced upon transition into the superconducting state \cite{morten04,morten05,poli08}. 

\begin{figure}[!tbh]
  \begin{minipage}[b]{\linewidth}
     \centerline{\includegraphics[clip=true,width=2.5in]{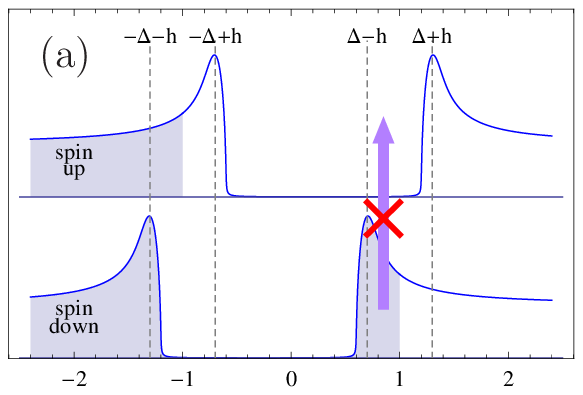}}
     \end{minipage}\hfill
   \begin{minipage}[b]{\linewidth}
   \centerline{\includegraphics[clip=true,width=2.5in]{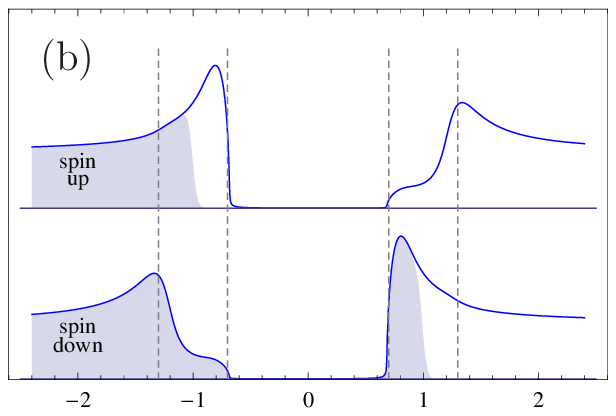}}
  \end{minipage}
\begin{minipage}[b]{\linewidth}
     \centerline{\includegraphics[clip=true,width=2.5in]{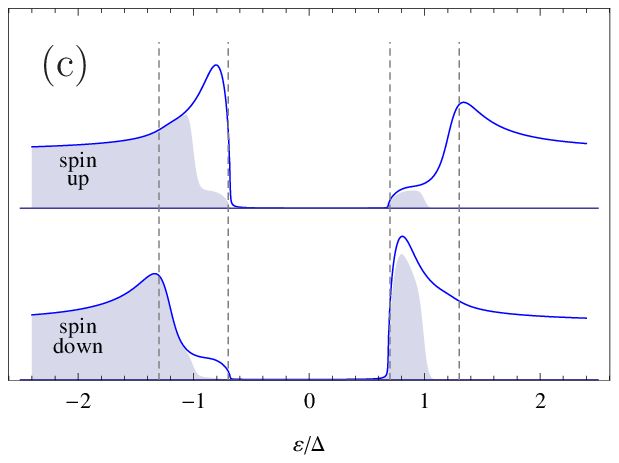}}
     \end{minipage}\hfill
   \caption{Schematic picture of the Zeeman-split superconducting DOS for the both spin subbands versus the quasiparticle energy. Its filling by the quasiparticles is also shown. (a) Only Zeeman splitting is taken into account upon calculating the DOS. There is a strict gap in the energy window of the Zeeman splitting for a one of spin subbands. There is a difference between spin-up and spin-dow distributions, but the elastic spin flips are blocked. (b) Spin-orbit scattering and magnetic impurities are taken into account upon calculating the DOS. The spin-flip processes are not blocked. The initial spin-dependent quasiparticle distribution is shown. (c) The same as in panel (b), but spin-up and spin-down distribution functions are already made equal by elastic spin flips.}   
\label{blocking}
\end{figure}

All the discussed above spin imbalance is caused by the difference in the distribution functions for spin-up and spin-down electrons. Therefore, the length of the corresponding spin relaxation is controlled by the length at which this difference disappears. At first glance, one can think that the Zeeman splitting of the superconducting DOS can greatly enhance the spin relaxation length. The idea is that in the energy window of the Zeeman splitting elastic spin flips cannot relax the nonequilibrium spin distribution function. This is because the processes are blocked by the absence of available DOS in one of spin subbands [see Fig.~\ref{blocking}(a)]. However, it was obtained \cite{bobkova15} that this mechanism does not work (at least for real values of elastic scattering rates). The reason is the following. The true energy gap is from $-\Delta+h$ to $\Delta-h$ and is the same for the both spin subbands, where $\Delta$ is the superconducting order parameter in the film and $h$ is the Zeeman field. There are no true energy gap in the energy window of the Zeeman splitting. Due to nonzero elastic scattering rates the DOS is redistributed between the spin subbands there [see Fig.~\ref{blocking}(b)]. Consequently, the elastic flips in the energy window of the Zeeman splitting are not blocked and the fast elastic spin-flip processes cancel the difference between spin-up and spin-down distributions in this energy window, as it is schematically shown in Fig.~\ref{blocking}(c).  As a result, the injected spin imbalance (we mean here the difference between the spin up and spin down distribution functions) relaxes in the Zeeman-split superconductor even faster than in the normal metal \cite{bobkova15}.

But inspecting Fig.~\ref{blocking}(c) one can suspect that the spin signal is still present in the superconductor in spite of the zero difference between spin-up and spin-down distributions. This is due to the different DOS in the spin subbands. Indeed, the quasiparticle current injected into the superconductor is accompanied by the energy flow that creates a nonequilibrium quasiparticle distribution in it, and the measured long-range spin signals were attributed to this spin-independent nonequilibrium quasiparticle distribution weighted by the spin-split DOS \cite{bobkova15,silaev15,krishtop15}. Such spin signals can be created even by normal (instead of ferromagnetic) injectors, as it was observed experimentally \cite{wolf13}. For such a spin signal a difference between distribution functions for quasiparticles with opposite spins is absent.  The relaxation length of such a spin signal is the energy relaxation length. The role of spin-flip processes is only to rapidly relax the distribution function to the spin-independent value. 

So, the origin of the measured signal the spin-independent difference between the distribution functions of electrons in the superconductor and in the detector. It simply can be converted into the spin signal by the spin-split DOS even if the spin-split DOS exists only near the detector. The long-range spin signal of the discussed type can be naturally divided into two parts according to the nature of this difference: due to nonthermalized quasiparticle distribution and due to thermalized overheated electron distribution. They can be distinguished by their different dependencies on the applied voltage. The first part of the signal is due to nonthermalized electrons. It is always present near the injector, where the injected high-energy electrons cannot be described by the Fermi distribution with a definite temperature. It is shown below that this part of the signal as a function of the injection voltage $V_I$ has typical one-peak shape, where the peak is located at $V_I \approx \Delta-h$. The main process providing thermalization is the electron-electron scattering. So, the decay length of this part of signal is determined by the electron-electron scattering rate, renormalized  due to superconductivity.
   
The second part of the signal is due to overheating of the injected electrons. It is determined by the difference between the effective electron temperature in the superconductor and in the detector. The typical shape of this signal is directly connected to the dependence of electron overheating temperature on the injection voltage: $T_e(V_I)$. In its turn, $T_e(V_I)$ is determined by superconducting density of states. That is, it has two steps at $V_I \approx \Delta \pm h$. Correspondingly, the measured nonlocal differential conductance $g_{non}=dI_D/dV_I$ manifests typical two-peak shape, where peaks are located at these voltages \cite{krishtop15}. The two peaks can be clearly observed in the signal if the coherent peaks in the superconducting DOS are well pronounced. It is worth to note here that this contribution can be viewed as a kind of a thermoelectric effect. The other types of giant thermoelectric effect were also predicted \cite{machon13,ozaeta14} and experimentally observed \cite{kolenda15} for superconductor/ferromagnet heterostructures .

The relaxation length of this thermalized contribution to the signal is controlled by the length, over which the effective electron temperature relaxes to its equilibrium value, that is the energy relaxation length. In dependence on the particular sample design it can be determined by the electron-phonon relaxation length or correspond to the distance between the injector and an equilibrium bulk reservoir. The latter case takes place if the heat leakage into the phonon subsystem can be neglected and the heat transport is controlled by the temperature gradient. Below it is shown that this scenario is more relevant to the existing low-temperature experiments with superconducting Al films. On the other hand, if one excludes the the heat leakage into the reservoir, the resulting relaxation length, controlled by the electron-phonon relaxation rate, can become very large.

The total measured nonlocal conductance has contributions from the both types of signal. Typically it manifests a pronounced peak at $V_I \approx \Delta - h$, provided by as thermalized, so as nonthermalized electrons. The second peak at $V_I \approx \Delta + h$ is only provided by thermalized electrons. It can be absent if the coherent peaks in the superconducting DOS are smeared enough by various depairing factors (first of all, by the orbital effect of the applied field) or if the electron overheating by the injected current is small, or if the detector is also overheated. It is also obvious that the peaks at $V_I \approx \Delta - h$ and at $V_I \approx \Delta + h$ should have different relaxation lengths. The former relaxes over the electron-electron relaxation length, while the relaxation of the latter is controlled by the length over which the electron subsystem is cooled.

It has been reported in the literature \cite{kopnin09,moor09} that the relaxation provided by the electron-electron scattering in Al at low temperatures is faster than the relaxation due to electron-phonon scattering. According to this reason in the present work we have clear hierarchy of length scales. The smallest length scales are the superconducting coherence length, elastic spin-flip length $\lambda_s$ and charge relaxation length (hundreds of nanometers). The larger scale is the electron-electron scattering length $\lambda_{e-e}$, renormalized due to superconductivity, applied magnetic field and electron overheating (it is of the order of several microns). And the largest scale is the length $L_h$ over which the electron subsystem is cooled. 

\begin{figure}[!tbh]
  \begin{minipage}[b]{\linewidth}
     \centerline{\includegraphics[clip=true,width=2.5in]{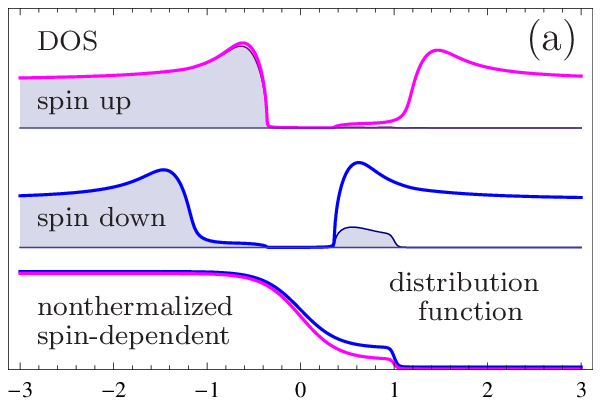}}
     \end{minipage}\hfill
    \begin{minipage}[b]{\linewidth}
   \centerline{\includegraphics[clip=true,width=2.5in]{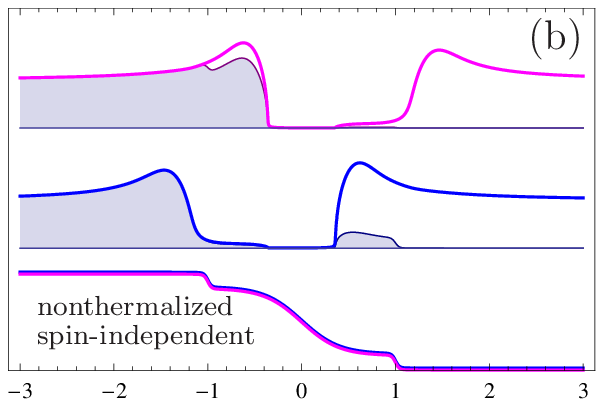}}
  \end{minipage}
\begin{minipage}[b]{\linewidth}
     \centerline{\includegraphics[clip=true,width=2.5in]{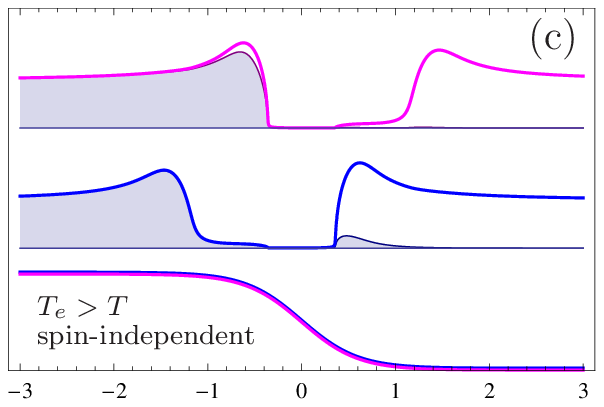}}
     \end{minipage}\hfill
    \begin{minipage}[b]{\linewidth}
   \centerline{\includegraphics[clip=true,width=2.5in]{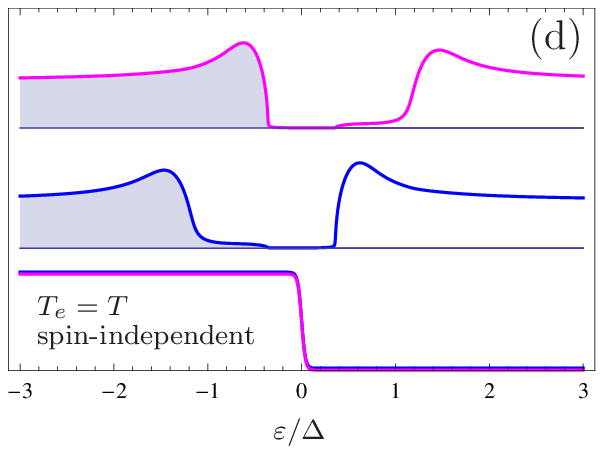}}
  \end{minipage}
   \caption{Space evolution of the distribution function, see text. The corresponding DOS, appropriately filled by quasiparticles, is also plotted.}   
\label{scales}
\end{figure}

The space evolution of the distribution function is schematically shown in Fig.~\ref{scales}. The distribution of injected electrons is spin-split [see Fig.~\ref{scales}(a)]. Also, it is nonthermalized. At distances $\sim \lambda_s$ it becomes spin-independent, as it is shown in panel (b) of Fig.~\ref{scales}. Now it is also symmetric over energy because the charge relaxation length has the same order. Further, at distances $\sim \lambda_{e-e}$ the distribution is thermalized, as in Fig.~\ref{scales}(c). Finally, the electrons are cooled at distances $\sim L_h$, as it is shown in panel (d).

Due to this hierarchy the second peak at $V_I \approx \Delta + h$ decays at larger distances than the first peak at $V_I \approx \Delta - h$. This conclusion is in agreement with the experimental results \cite{wolf14}. 

\section{theoretical approach}

\label{approach}

\subsection{basic equations}

As it was already mentioned above, we focus on the nonequilibrium spin accumulation $S$. This quantity can be written in terms of the Keldysh quasiclassical Green function as $S=-\int \limits_{-\infty}^\infty d\varepsilon {\rm Tr} \left[\tau_3 \sigma_3 \left(\check g^K-\check g^K_{eq}\right)\right] /16$, where $\tau_i$ and $\sigma_i$ are Pauli matrices in the particle-hole and spin spaces, respectively. $\check g^K$ is the Keldysh component ($4 \times 4$ matrix) of the quasiclassical Green's function $\check g = \left( \begin{array}{cc} \check g^R & \check g^K \\ 0 & \check g^A \end{array} \right)$, where $\check g^{R(A)}$ are retarded and advanced Green's functions. $\check g^K_{eq}$ means the value of the Keldysh component in equilibrium. We assume the superconductor to be in the diffusive limit, so the matrix $\check g$ obeys the Usadel equation \cite{usadel,bergeret05}
\begin{eqnarray}
D \hat \partial_y (\check g \hat \partial_y \check g)\! +\! i\! \left[ \check \Lambda-\check \Sigma_{so}-\check \Sigma_{mi}-\check \Sigma_{e-e}-\check \Sigma_{e-ph},\check g \right]=0.~~~~
\label{usadel}
\end{eqnarray}
Here $\check \Lambda = \varepsilon \tau_3- h \sigma_3 \tau_3-\Delta i\tau_2$, $\varepsilon$ is the quasiparticle energy, $D$ is the diffusion 
constant, and $h=\mu_B H+h_{int}(H)$ is the Zeeman field.  Here we assume that there are two sources of the Zeeman field in the film: the first contribution is caused by the applied magnetic field $H$ and the second contribution is the effective internal exchange field $h_{int}(H)$, which can be induced in the superconducting film if it is fabricated on top of the ferromagnetic insulator. In real experimental situation the internal  exchange field depends on the applied magnetic field \cite{wolf14}. In our study we take the simple phenomenological model $h_{int}(H)=0.4 \Delta_0 \tanh[\mu_B H/0.08\Delta_0]$ for this dependence. $\Delta_0$ is the zero-temperature superconducting order parameter in the absence of the magnetic field and the ferromagnetic insulator. This model corresponds qualitatively to the experimental one \cite{wolf14}, and the exact law $h_{int}(H)$ does not influence qualitatively our results.  

$\hat \partial_y$ is a matrix in particle-hole space, accounting for the orbital suppression of superconductivity by the magnetic field. 
For a general matrix $\check G$ in particle-hole space $\hat \partial_y \check G =\partial_y \check G -(ie/c)(Hx+A_0)\left[ \tau_3, \check G \right]$, where $x$ is the coordinate normal to the film. Eq.~(\ref{usadel}) should be supplemented by the normalization condition $\check g^2=1$.

The terms $\check \Sigma_{so}=\tau_{so}^{-1}(\bm \sigma \check g \bm \sigma)$ and $\check \Sigma_{mi}=\tau_{mi}^{-1}(\bm \sigma \tau_3 \check g \bm \sigma \tau_3)$ in Eq.~(\ref{usadel}) describe elastic spin relaxation processes of spin-orbit scattering and exchange interaction with magnetic impurities, respectively. The last terms $\check \Sigma_{e-e}$ and $\check \Sigma_{e-ph}$ describe electron-electron and electron-phonon relaxation, respectively.

We assume that the transparencies of the injector and detector interfaces are small, so that up to the leading (zero) order in transparency the retarded, advanced Green's functions and the order parameter take their bulk values. The Green's functions can be represented in the form $\check g^R=g_0^R \tau_3 + g_t^R \sigma_3 \tau_3 + f_0^R i\tau_2 +f_t^R \sigma_3 i \tau_2$. It is convenient to use the following $\theta$-parametrization, which satisfies the normalization condition: $g_{0,t}^R=(\cosh \theta_+ \pm \cosh \theta_-)/2$ and $f_{0,t}^R=(\sinh \theta_+ 
\pm \sinh \theta_-)/2$. The advanced Green's functions can be found as $\check g^A=-\check g^{R*}$. 

We assume that the film thickness in the $x$ direction is smaller than the superconducting coherence length. Then $\theta_\pm$ does not depend on $x$. Integrating the retarded part of Eq.~(\ref{usadel}) over the width $d$ of the film along the $x$-direction, one can obtain from Eq.~(\ref{usadel}) that $\theta_{\pm}$ obeys the following equation:
\begin{eqnarray}
(\varepsilon \mp h)\sinh \theta_\pm +\Delta \cosh \theta_\pm +  \nonumber \\
Di \frac{e^2}{6c^2}H^2d^2 \cosh \theta_\pm \sinh \theta_\pm \pm 2i \tau_{so}^{-1}\sinh(\theta_+ - \theta_-)+\nonumber \\
2i\tau_{mi}^{-1}[\cosh \theta_\pm \sinh \theta_\pm +\sinh (\theta_+ +\theta_-)]=0
\label{theta}
\enspace .~~~~
\end {eqnarray}

Here the third term describes the orbital depairing of superconductivity. Usually this orbital deparing can be disregarded for thin films in parallel magnetic field. However, it can be estimated that for magnetic fields of the order of 1-2 T, which are applied in experiment, the orbital depairing can even exceed the other depairing factors (spin-orbit and magnetic impurity scattering). So, it cannot be neglected in Eq.~(\ref{theta}). $\Delta$ is calculated self-consistently taking into account its suppression by the applied field, internal exchange field, spin-orbit deparing and deparing by magnetic impurities. The corresponding self-consistency equation is as follows:
\begin{equation}
\Delta=\int \limits_{-\omega_D}^{\omega_D} \frac{d\varepsilon}{4}\Lambda \sum \limits_\sigma {\rm Re} \sinh \theta_\sigma \tanh \frac{\varepsilon}{2T}.
\label{self-consistency}
\end{equation}
Here $\Lambda$ is the dimensionless coupling constant and $\omega_D$ is the high energy cut-off. The suppression of the order parameter due to nonequilibrium quasiparticle distribution in the film is not taken into account by Eq.~(\ref{self-consistency}) and the distribution function is taken equal to its equilibrium value. As it is shown below, the distribution function in the film can be represented as $\varphi(\varepsilon)=\tanh (\varepsilon /2T_e)+\delta \varphi_\varepsilon$. When the transparency of the I/S interface is small, the second term in this expression is of the first order in this transparency and, therefore, can be disregarded up to the leading order. However, the electron overheating temperature is a non-analytic function of the transparency. So, our assumption of the equilibrium distribution function in Eq.~(\ref{self-consistency}) is strictly valid only for low enough electron overheating. When the electron overheating rises, Eq.~(\ref{self-consistency}) underestimates the suppression of the order parameter.  

The terms $\check \Sigma_{e-e}$ and $\check \Sigma_{e-ph}$, in principle, also enter Eq.~(\ref{theta}) as another depairing factor, but it is neglected because at low temperature it is small as compared to other depairing factors. It is important only for the calculation of the distribution function.

The normalization condition allows to write the Keldysh component as $\check g^K=\check g^R \check \varphi -\check \varphi \check g^A$, where $\check \varphi$ is the distribution function with the following general structure in particle-hole and spin spaces: $\check \varphi = (1/2)[\varphi_+^0 + \varphi_+^t\sigma_z + \varphi_-^0 \tau_z + \varphi_-^t \tau_z \sigma_z]$. Physically the distribution function $\varphi_-$ is responsible for the charge imbalance and $\varphi_+$ for the spin imbalance in the system. The components $\varphi_\pm^0$ describe the spin-independent part of the quasiparticle distribution, while $\varphi_\pm^t$ accounts for its spin polarization. In the equilibrium $\varphi_+^{0,eq}=2 \tanh (\varepsilon/2T)$ and the other components of $\check \varphi$ are zero. Via the distribution function the nonequilibrium spin accumulation $S$ can be written as follows
\begin{equation}
S=-\frac{1}{4} \int \limits_{-\infty}^\infty d \varepsilon \left( {\rm Re}[g_t^R](\varphi_+^0-2\tanh \frac{\varepsilon}{2T}) + {\rm Re}[g_0^R]\varphi_+^t \right)  
\label{spin}
\enspace .
\end{equation}    
It is worth to note here that for Zeeman-split superconductor the triplet part of the normal Green's function $g_t^R $ is nonzero, while it is vanishes for $h=0$. Due to this fact the nonequilibrium spin accumulation $S$ can be nonzero in the Zeeman-split superconductor even for the case of spin-independent quasiparticle distribution, that is for $\varphi_+^t=0$. In principle, for a Zeeman-split superconductor there is an equilibrium spin accumulation near the Fermi energy $S_{eq}=-\frac{1}{2} \int \limits_{-\infty}^\infty d \varepsilon  {\rm Re}[g_t^R]\tanh (\varepsilon/2T) $. But we do not consider this quantity here because it does not contribute to the measured signal.

The equations for the distribution functions $\varphi_+^{0,t}$, entering Eq.~(\ref{spin}), can be derived from Eq.~(\ref{usadel}) and take the form
\begin{eqnarray}
D(\kappa_1 \partial_y^2 \varphi_+^0 + \kappa_2 \partial_y^2 \varphi_+^t)- \nonumber \\ 
\frac{(I_{\uparrow,e-e}+I_{\downarrow,e-e})}{2}-\frac{(I_{\uparrow,e-ph}+I_{\downarrow,e-ph})}{2}=0, \label{varphi1} \\
D(\kappa_2 \partial_y^2 \varphi_+^0 + \kappa_1 \partial_y^2 \varphi_+^t)-K\varphi_+^t- \nonumber \\ 
\frac{(I_{\uparrow,e-e}-I_{\downarrow,e-e})}{2}-\frac{(I_{\uparrow,e-ph}-I_{\downarrow,e-ph})}{2}=0
\label{varphi2}
\enspace .
\end{eqnarray} 
Here $\kappa_1=1+|g_0^R|^2+|g_t^R|^2-|f_0^R|^2-|f_t^R|^2$ and $\kappa_2=2{\rm Re}[g_0^R g_t^{R*}-f_0^R f_t^{R*}]$ account for the renormalization of the diffusion constant by superconductivity. $K=K_{so}+K_{mi}$ is responsible for the spin relaxation by elastic processes: spin-orbit scattering and spin-flip scattering by magnetic impurities, and
\begin{eqnarray}
K_{so(mi)}=8\tau_{so(mi)}^{-1}[ {\rm Re}({g_0^R}^2\mp {f_0^R}^2)+|g_0^R|^2\mp |f_0^R|^2- \nonumber \\
{\rm Re}({g_t^R}^2\mp {f_t^R}^2)-(|g_t^R|^2\mp |f_t^R|^2) ]
\label{K}
\enspace .~~~~~
\end{eqnarray}
The collision integrals $I_{\sigma,e-e}$ and $I_{\sigma,e-ph}$ in Eqs.~(\ref{varphi1})-(\ref{varphi2}) describe electron-electron and electron-phonon relaxation processes, respectively.

Kinetic equations Eqs.~(\ref{varphi1})-(\ref{varphi2}) shoud be applied by the appropriate boundary conditions at the injector/superconductor interface. These boundary conditions are to be obtained from the general Kupriyanov-Lukichev boundary conditions \cite{kuprianov88}, generalized for spin-filtering interfaces \cite{bergeret12,machon14}. In the considered case up to the leading order in the junction transparency we can neglect the superconducting proximity effect in the injector electrode. In this case the spectral function in it has a trivial spin and particle-hole structure: $\check g_I^{R,A}=\pm \tau_3$. Then the boundary conditions take the form   
\begin{equation}
\check g \hat \partial_y \check g = - \frac{\check G}{2\sigma_s}\left[ \check g, \check g_I \right]
\label{KL}
\enspace .
\end{equation}
If the injector is biased with respect to the superconductor by the voltage $V_I$, the Keldysh Green's function there takes the form $\check g_I^K=\tau_3 (\varphi_{I+}^0+\varphi_{I-}^0\tau_3)$, where
$\varphi_{I\pm}^0=\tanh[(\varepsilon-V_I)/2T]\pm\tanh[(\varepsilon+V_I)/2T]$. The tunnel interface between the injector and the superconductor is assumed to be spin-polarized with the conductance matrix $\check G=G_0+G_t\tau_3\sigma_3$. $\sigma_s$ is the conductivity of the superconductor. 

In the tunnel limit we consider the injected current polarization is mainly determined by the spin polarization of the tunnel conductance $P_I=G_t/G_0$. While in experiment as ferromagnetic, so as normal injectors were used, for simplicity we consider only normal injectors with $P_I=0$ in the present work. The results for the ferromagnetic injectors are qualitatively the same, the only difference is that the nonlocal conductance shape disturbs slightly from the purely antisymmetric form \cite{bobkova15,silaev15}. This is in agreement with the experimental results \cite{hubler12,wolf13}. 

Boundary conditions for the distribution functions at $y=0$ are to be obtained making use of the Keldysh part of Eq.(\ref{KL}). They take the form:
\begin{widetext}
\begin{eqnarray}
\kappa_1 \partial_y \varphi_+^0+\kappa_2 \partial_y \varphi_+^t+\frac{2G_0}{\sigma_s}\left\{[{\rm Re}g_0^R](\varphi_{I+}^0-\varphi_+^0)-[{\rm Re}g_t^R]\varphi_+^t\right\}+ \frac{2G_t}{\sigma_s}\left\{[{\rm Re}g_t^R](\varphi_{I-}^0-\varphi_-^0)-[{\rm Re}g_0^R]\varphi_-^t\right\}=0.\label{varphi_bc1}\\
\kappa_1 \partial_y \varphi_+^t+\kappa_2 \partial_y \varphi_+^0+\frac{2G_0}{\sigma_s}\left\{[{\rm Re}g_t^R](\varphi_{I+}^0-\varphi_+^0)-[{\rm Re}g_0^R]\varphi_+^t\right\} + \frac{2G_t}{\sigma_s}\left\{[{\rm Re}g_0^R](\varphi_{I-}^0-\varphi_-^0)-[{\rm Re}g_t^R]\varphi_-^t\right\}=0 
\label{varphi_bc2}
\end{eqnarray}
\end{widetext} 
It is worth to note here that, while the distribution functions $\varphi_+$ and $\varphi_-$ obey the independent kinetic equations, they are coupled by the boundary conditions, if the interface barrier is spin-polarized, as it is seen from Eqs.~(\ref{varphi_bc1})-(\ref{varphi_bc2}).

As it was already mentioned above, we assume that the elastic spin-flip processes are much faster than the electron-electron and electron-phonon relaxation, that is there is a small parameter $\tau_\varepsilon^{-1}/K \ll 1$ in the considered problem. Here $\tau_\varepsilon^{-1}$ is the characteristic scattering rate of the e-e and e-ph relaxation. This assumption is in good agreement with the experimental situation \cite{hubler12,quay13}.  

Under this condition the solution of Eqs.~(\ref{varphi1})-(\ref{varphi2}) takes the form
\begin{eqnarray}
\left(\begin{array}{c} 
\varphi_+^0 \\
\varphi_+^t \end{array}\right)=\alpha 
\left(\begin{array}{c} 
-\frac{\kappa_2}{\kappa_1} \\
 1 \end{array}
\right)e^{-\lambda_s y}+
\left(\begin{array}{c}\tilde \varphi_+^0 (y) \\
\tilde \varphi_+^t(y) \end{array}\right)
\label{varphi_sol}
\enspace ,~~~~
\end{eqnarray}
where  $\lambda_s^2=\kappa_1 K/D(\kappa_1^2-\kappa_2^2)$. The first term in Eq.~(\ref{varphi_sol}) describes fast spin relaxation of the distribution function due to elastic spin-flip processes. The second term corresponds to slow e-e and e-ph relaxation to the equilibrium form. This term is spin-independent up to the leading order in the small parameter $\tau_\varepsilon^{-1}/K $, that is $\tilde \varphi_+^t \sim (\tau_\varepsilon^{-1}/K) \tilde \varphi_+^0$. 
We would like to stress here the role of the elastic spin flips in the considered problem. Although the corresponding characteristic length is small, the elastic spin flips qualitatively influence the results at any distances from the injector when the electron distribution is nonequilibrium. The point is that the superconducting DOS is spin-dependent. Under this condition the electron-electron and electron-phonon scattering makes the electron distributions for spin up and spin down electrons to be different even if the injected electron distribution is spin-independent. It is the elastic spin flips that average the spins providing the spin-independent electron distribution.

Further we are interested in the slow e-e and e-ph relaxation of this approximately spin-independent distribution function. The next subsection is devoted to the electron-electron collision integral. 

\subsection{electron-electron relaxation}

Being far from the Stoner instability we neglect the triplet channel for the electron-electron interaction. In this case there are two possible processes allowed: (i) in the initial state quasiparticles have the same spin projections on the quantization axis, which remain unchanged during the collision  and  (ii) in the initial state quasiparticles have different spin projections, which again remain unchanged during the collision. Each process corresponds to a certain term in the collision integral:
\begin{eqnarray}
J_\sigma^{(1)}=\int d \varepsilon' N_\sigma (\varepsilon) N_\sigma (\varepsilon+\omega) N_\sigma(\varepsilon') N_\sigma(\varepsilon'+\omega) \left[ (\varphi_{\varepsilon+\omega}- \right.\nonumber \\
\left. \varphi_{\varepsilon})(1-\varphi_{\varepsilon'}\varphi_{\varepsilon'+\omega})-(\varphi_{\varepsilon'+\omega}-\varphi_{\varepsilon'})(1-\varphi_{\varepsilon} \varphi_{\varepsilon+\omega}) \right],~~~~~~
\label{phase_1}
\end{eqnarray}
\begin{eqnarray}
J_\sigma^{(2)}=\int d \varepsilon' N_\sigma (\varepsilon) N_\sigma (\varepsilon+\omega) N_{\bar \sigma}(\varepsilon') N_{\bar \sigma}(\varepsilon'+\omega)\left[ (\varphi_{\varepsilon+\omega}-\right. \nonumber \\
\left.\varphi_{\varepsilon})(1-\varphi_{\varepsilon'}\varphi_{\varepsilon'+\omega})-(\varphi_{\varepsilon'+\omega}-\varphi_{\varepsilon'})(1-\varphi_{\varepsilon}\varphi_{\varepsilon+\omega}) \right],~~~~~~
\label{phase_2}
\end{eqnarray}
where $N_\sigma(\varepsilon)$ is the superconducting DOS normalized to the normal-state DOS $N_F$ at the Fermi level. It can be obtained from the retarded part of the Green's function in a standard way. $\varphi_\varepsilon \equiv \varphi_+^0(\varepsilon)$ is the distribution function. It is assumed to be spin-independent here according to the disscussed above. In equilibrium $\varphi_{+,eq}(\varepsilon)=2\tanh(\varepsilon)/2T$.

The two-quasiparticle collision integral can be represented as: \cite{dimitrova07}
\begin{equation}
I_{\sigma,e-e}(\varepsilon)=\sum \limits_{p=1,2}\int \frac{d \omega}{2\pi N_F}d\varepsilon'K_p(\omega)J_\sigma^{(p)}(\varepsilon,\omega)
\label{ee_general}
\enspace ,
\end{equation}
where $K_p(\omega)$  describes the strength of relaxation due to the corresponding processes. First of all, these quantities do not depend on spin, because they correspond to the singlet processes. Second, $K_1(\omega)=K_2(\omega)$ because we neglect the Fermi-liquid constant corresponding to the triplet interaction channel. In addition, for simplicity we assume that the interaction region is shorter than the mean free path $l$, so for the collision integral the quasiparticle dynamics can be considered as ballistic. In this case the kernel $K$ does not depend on $\omega$ \cite{altshuler}. However, all the results can be generalized for the disordered quasiparticle dynamics as well. 

To proceed further we linearize the collision integral with respect to the deviation of the distribution function from its thermalized value $\delta \varphi_+ (\varepsilon)=\varphi_+(\varepsilon)-2\tanh(\varepsilon/2T_e)$. Here $T_e\equiv T_e(V_I,y) > T$ is the effective temperature of the electronic subsystem. The resulting expressions are as follows:
\begin{widetext}
\begin{eqnarray}
J_\sigma^{(1)}=\int d \varepsilon' N_\sigma (\varepsilon+\omega) N_\sigma(\varepsilon') N_\sigma(\varepsilon'+\omega) N_\sigma(\varepsilon)
\left\{ \delta \varphi_{\varepsilon} \left[ \tanh \frac{\varepsilon+\omega}{2T_e}(\tanh \frac{\varepsilon'+\omega}{2T_e}-\tanh \frac{\varepsilon'}{2T_e})-1+\tanh \frac{\varepsilon'}{2T_e}\tanh \frac{\varepsilon'+\omega}{2T_e} \right]+\right.\nonumber \\
\delta \varphi_{\varepsilon+\omega} \left[ \tanh \frac{\varepsilon}{2T_e}(\tanh \frac{\varepsilon'+\omega}{2T_e}-\tanh \frac{\varepsilon'}{2T_e})+1-\tanh \frac{\varepsilon'}{2T_e}\tanh \frac{\varepsilon'+\omega}{2T_e} \right]+ \nonumber ~~~~~~~~~~~~~~~~~~~~\\
\delta \varphi_{\varepsilon'} \left[ \tanh \frac{\varepsilon'+\omega}{2T_e}(\tanh \frac{\varepsilon}{2T_e}-\tanh \frac{\varepsilon+\omega}{2T_e})+1-\tanh \frac{\varepsilon}{2T_e}\tanh \frac{\varepsilon+\omega}{2T_e} \right]+ \nonumber ~~~~~~~~~~~~~~~~~~~~\\
\left. \delta \varphi_{\varepsilon'+\omega} \left[ \tanh \frac{\varepsilon'}{2T_e}(\tanh \frac{\varepsilon}{2T_e}-\tanh \frac{\varepsilon+\omega}{2T_e})-1+\tanh \frac{\varepsilon}{2T_e}\tanh \frac{\varepsilon+\omega}{2T_e} \right] \right \},~~~~~~~~~~~~~~~~~~~~
\label{phase_1}
\end{eqnarray}
\end{widetext}
$J_\sigma^{(2)}$ can be obtained from Eq.~(\ref{phase_1}) by substitution $N_\sigma(\varepsilon') N_\sigma(\varepsilon'+\omega) \to N_{\bar \sigma}(\varepsilon') N_{\bar \sigma}(\varepsilon'+\omega)$.

Our final goal is to calculate the nonlocal conductance $g_{nl}=dI_D/dV_I$. For this purpose we only need the derivative $d\varphi_+^0/dV_I$, according to Eq.~(\ref{spin}). At low injector temperatures the derivative $d\delta \varphi_\varepsilon/dV_I$ can be divided into the singular and regular contributions:
\begin{equation}
d\delta \varphi_\varepsilon/dV_I = \tilde \varphi_{1} (V_I)\delta (\varepsilon-V_I)+\tilde \varphi_{2} (-V_I)\delta (\varepsilon+V_I)+\varphi_{reg}.
\label{sing_reg}
\end{equation}
For the singular parts the electron-electron collision integral in the kinetic equation becomes local by energy, or, in other words, it can be treated in the $\tau-$approximation. The corresponding inverse relaxation time can be found from Eqs.~(\ref{ee_general})-(\ref{phase_1}) and takes the form
\begin{eqnarray}
\tau_{e-e,\sigma}^{-1}=-\frac{\gamma_{e-e}(T_c)}{4T_c^2}\int d\omega d\varepsilon'N_\sigma(\varepsilon+\omega)N_\sigma(\varepsilon)\times \nonumber \\
\left( N_\sigma(\varepsilon')N_\sigma(\varepsilon'+\omega)+N_{\bar \sigma}(\varepsilon')N_{\bar \sigma}(\varepsilon'+\omega) \right)\times \nonumber \\
\left[ \tanh \frac{\varepsilon+\omega}{2T_e}(\tanh \frac{\varepsilon'+\omega}{2T_e}-\tanh \frac{\varepsilon'}{2T_e})-\right. \nonumber \\
\left.1+\tanh \frac{\varepsilon'}{2T_e}\tanh \frac{\varepsilon'+\omega}{2T_e} \right].~~~~~~~~~~~
\label{tau_ee}
\end{eqnarray}
For a normal metal limit $N_\sigma(\varepsilon) \to 1$ and at low temperatures Eq.~(\ref{tau_ee}) takes the well-known form $\tau_{e-e}^{-1}=(\gamma_{e-e}(T_c)/T_c^2)\varepsilon^2$. In our calculations we assume $\tau_{e-e}(T_c)=1$ns.

We assume that the contribution of the regular part of the nonthermalized quasiparticle distribution $\varphi_{reg}$ into the nonlocal conductance can be neglected as compared to the contribution of the singular part. In this case the total electron-electron collision integral in Eq.~(\ref{varphi1}) can be treated in the $\tau$-approximation as follows 
\begin{equation}
I_{e-e,\sigma}(\varepsilon)=\frac{\delta \varphi_\varepsilon}{\tau_{e-e,\sigma}(\varepsilon)}
\label{tau_approx}
\end{equation} 
with $\tau_{e-e,\sigma}^{-1}$ determined by Eq.~(\ref{tau_ee}). 

For the considered problem the discussed above assumption of "dominating singular part" can be violated in two cases: (i) for small voltages $V_I$  less than the gap value. However, the nonlocal conductance is practically zero in this region due to the absence of nonzero DOS, so this voltage region is not essential; (ii) if the electron temperature $T_e$ in the film depends strongly on $V_I$. In this case our $\tau$-approximation can give only qualitative results.

Typical dependence of $\tau_{e-e,\sigma}^{-1}$, determined by Eq.~(\ref{tau_ee}), on quasiparticle energy is $\sim \varepsilon^2$ (like in the normal metal), but this is valid starting from $\varepsilon \sim 3\varepsilon_g$, where $\varepsilon_g$ is the spectral gap. It is natural, because in order for an injected electron to be scattered by an equilibrium electron, the former should have an energy more than $3\varepsilon_g$. For energies less than $3\varepsilon_g$ $\tau_{e-e,\sigma}^{-1}$ decreases exponentially when the temperature goes down.   

\subsection{electron overheating and relaxation of the electron temperature}

\label{phonon}

The next our goal is to calculate the electron overheating temperature $T_e(y)$. First of all, $T_e(y=0)$ enters Eq.~(\ref{tau_ee}) for the inverse electron-electron relaxation time, therefore it affects crucially the relaxation length of the nonthermalized part of the spin signal. Second, the difference $T_e(y)-T$ determine the thermal part of the signal.

$T_e(y)$ should be calculated from the heat balance equation. This equation can be obtained by multiplying the kinetic equation Eq.~(\ref{varphi1}) by $\varepsilon$ and integrating over the quasiparticle energy and over the $y$-coordinate. The contribution of the electron-electron collision term equals zero because this term
conserves the total energy. Then the heat balance equation takes the form
\begin{eqnarray}
\eta  \Phi_I=\Phi(y)-\int \limits_0^y dy \int \limits_{-\infty}^\infty \varepsilon d\varepsilon  \frac{I_{e-ph,\uparrow }+I_{e-ph,\downarrow }}{2},
\label{heat_balance}
\end{eqnarray}
where 
\begin{equation}
\Phi(y)=D \int \limits_{-\infty}^\infty \varepsilon d\varepsilon \left[ \kappa_1 \partial_y \varphi_+^0 + \kappa_2 \partial_y \varphi_+^t \right]
\label{heat_current}
\end{equation} 
is the heat current at distance $y$ from the injector and $\Phi_I=\Phi(y=0)$ is the heat current injected at $y=0$. This injected heat current can be calculated from the boundary condition Eq.~(\ref{varphi_bc1}). $0<\eta<1$ is a dimensionless phenomenological parameter, accounting for the fact that only a fraction of the injected heat travels along the Al strip towards the detector. In case if the system is symmetric with respect to injector point and there is no reverse heat leakage into the injector $\eta=1/2$. The second term in the r.h.s of Eq.~(\ref{heat_balance}) describes the heat leakage into the phonon subsystem.

Eq.~(\ref{heat_balance}) should be supplied by the boundary condition, which depends on the particular model of electron cooling. One can  assume that all the injected heat goes into the phonon subsystem. In this case the boundary condition is $T_e(y) \to T$ at $y \to \infty$. The other possible realistic situation is that the superconducting strip is attached to a massive reservoir at a distance $L_h$ from the injector, so there is a heat current into the reservoir. Also the heat can leak into the detector electrode. Of course, in real setup all these channels of heat leakage contribute, but we assume the transparency of the interface between the film and the detector to be small, so the corresponding heat current can be disregarded. 

First of all, we estimate the characteristic length of electron cooling due to electron-phonon relaxation. The corresponding electron-phonon collision integral can be obtained from the general expression in terms of the quasiclassical Green's functions  \cite{kopnin}. For our kinetic equation, which is written for the distribution function $\varphi_+$, it takes the form
\begin{eqnarray}
I_{e-ph,\sigma}=\frac{\gamma_{e-ph}}{2T_c^3} \int d \varepsilon' (\varepsilon'-\varepsilon)^2 {\rm sign} (\varepsilon'-\varepsilon)  \times ~~~~~~\nonumber \\ 
4\left[ \coth \frac{\varepsilon'-\varepsilon}{2T}(\varphi_+(\varepsilon)-\varphi_+(\varepsilon'))-\frac{1}{2}\varphi_+(\varepsilon) \varphi_+(\varepsilon')+2  \right] \times \nonumber \\
 \left[ {\rm Re}g_\sigma^R(\varepsilon){\rm Re}g_\sigma^R(\varepsilon')-{\rm Re}f_\sigma^R(\varepsilon){\rm Re}f_\sigma^R(\varepsilon') \right]~~~~~~~~~~
\label{e-ph_full}
\end{eqnarray}
At distances $y \gg \lambda_{e-e}$, which are of interest for the electron-phonon relaxation, the distribution function is already thermalized: $\varphi_+ \approx 2\tanh[{\varepsilon}/{2T_e}]$. In principle, the distribution function $\varphi_-$ also enters the electron-phonon collision integral, but it disappears on the charge relaxation length, which is even much smaller  than $\lambda_{e-e}$, so it can be omitted in Eq.~(\ref{e-ph_full}). 
In normal state at low temperatures the linearized (with respect to $\varphi_+-2\tanh[{\varepsilon}/{2T}]$) version of Eq.~(\ref{e-ph_full}) gives well-known answer $\tau_{e-ph,n}^{-1}=(\gamma_{e-ph}/T_c^3)\varepsilon^3$. In our calculation we choose $\tau_{e-ph}(T_c)=100$ns. It corresponds to the normal state electron-phonon relaxation length $\lambda^N_{e-ph}\approx 180\xi_0$, where $\xi_0=\sqrt{D/\Delta_0}$ is the zero-temperature coherence length in the considered superconducting film. Typically $\xi_0$ is of the order of $100-200$nm in superconducting Al, so $\lambda_N$ is of order of $20-30$ microns.   

\begin{figure}[!tbh]
  \centerline{\includegraphics[clip=true,width=3.0in]{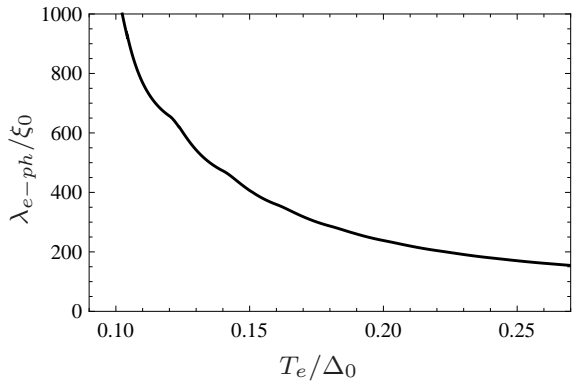}}
  \caption{$\lambda_{e-ph}$ vs the electron temperature $T_e$ at $y=0$. $\tau_{so}^{-1}=\tau_{mi}^{-1}=0.015\Delta_0$, $\mu_B H=0.10 \Delta_0$, $h_{int=0}$. Here and below throughout the paper the temperature of the reservoirs $T=0.02\Delta_0$.}   
\label{e-ph_length}
\end{figure}
 
Making use of Eqs.~(\ref{heat_balance}), (\ref{e-ph_full}) and the boundary condition $T_e(y) \to T$ at $y \to \infty$ we calculate numerically the corresponding length $\lambda_{e-ph}$ of electron cooling. It is represented in Fig.~\ref{e-ph_length} versus $T_{e,0} \equiv T_e(y=0)$. It is seen that due to the superconducting renormalization this length is typically of the order of hundreds of microns. For this reason in our further calculation we consider the other model of electron cooling, where the superconducting strip is attached to a massive reservoir at a distance $\lambda_{e-e} \ll \L_h \ll \lambda_{e-ph}$ from the injector. This model is technically much simpler and allows for neglecting the heat leakage into the phonon subsystem.

In this case one can omit $I_{e-ph}$ term in Eq.~(\ref{heat_balance}) and the heat current approximately conserves: $\Phi(y)=const$. Then the electron overheating temperature $T_e(y)$ can be obtained from the following equation:
\begin{equation}
2D \int \limits_{-\infty}^{\infty} \varepsilon d\varepsilon \kappa_1(\varepsilon) \left[ \tanh \frac{\varepsilon}{2T_e}-\tanh \frac{\varepsilon}{2T} \right]=\eta (y-L_h) \Phi_I .
\label{Te}
\end{equation}

\subsection{calculation of the distribution function}

As it was already discussed in Sec.~\ref{model}, the long-range spin signal can be divided into two physically different contributions: due to nonthermalized nonequilibrium quasiparticles $S_{nth}$ and due to the thermalized overheated quasiparticles $S_{th}$. The resulting expressions are as follows:
\begin{eqnarray}
S_{nth}=-\frac{1}{4} \int \limits_{-\infty}^\infty d \varepsilon {\rm Re}[g_t^R]\delta \varphi_\varepsilon,~~~~~~~~~~~ 
\label{nontherm_spin} \\
S_{th}=-\frac{1}{2} \int \limits_{-\infty}^\infty d \varepsilon {\rm Re}[g_t^R]\left( \tanh \frac{\varepsilon}{2T_e}-\tanh \frac{\varepsilon}{2T}\right).
\label{therm_spin}
\end{eqnarray}
While the contribution $S_{th}$ can be calculated directly from Eq.~(\ref{therm_spin}) having at hand the electron overheating temperature $T_e$, in order to obtain $S_{nth}$ one should at first calculate the nonthermalized part of the distribution function $\delta \varphi_\varepsilon$. It should be found from the kinetic equations Eqs.~(\ref{varphi1})-(\ref{varphi2}), where the electron-phonon relaxation term is omitted and the electron-electron relaxation term is taken in the $\tau$-approximation Eq.~(\ref{tau_approx}).

Under these conditions the solution of Eqs.~(\ref{varphi1})-(\ref{varphi2}) up to the leading order in the parameter $\tau_{e-e}^{-1}/K $ takes the form
\begin{eqnarray}
\left(\begin{array}{c} 
\delta\varphi_\varepsilon \\
\varphi_+^t \end{array}\right)=\alpha \left(\begin{array}{c} 
-\frac{\kappa_2}{\kappa_1} \\
 1 \end{array}
\right)e^{-\lambda_s y}+
\beta \left(\begin{array}{c} 1 \\
\frac{\tau_{e-e}^{-1}\kappa_2}{K\kappa_1} \end{array}\right)e^{-\lambda_{e-e} y} 
\label{varphi_sol_ee}
\enspace ,~~~~
\end{eqnarray}
where $\delta \varphi_\varepsilon =\varphi_+^0 -2\tanh \frac{\varepsilon}{2T_e}$, $\tau_{e-e}^{-1}=(1/2)(\tau_{e-e,\uparrow }^{-1}+\tau_{e-e,\downarrow }^{-1})$, and $\lambda_{e-e}^2=\tau_{e-e}^{-1}/D\kappa_1$. The first term in Eq.~(\ref{varphi_sol_ee}) coincides with the first term in Eq.~(\ref{varphi_sol}) and describes fast spin relaxation of the distribution function due to elastic spin-flip processes. The second term is of interest now and gives the nonthermalized part of the distribution function. As it was discussed above, its spin part $\varphi_+^t$ is small and its spin-independent part $\delta \varphi_\varepsilon$ should be substituted into Eq.~(\ref{nontherm_spin}).

Constants $\alpha$ and $\beta$ should be found from the boundary conditions Eqs.~(\ref{varphi_bc1})-(\ref{varphi_bc2}) at the injector/superconductor interface. Up to the leading order in $\tau_\varepsilon^{-1}/K$ and up to the leading order in the I/S interface conductance it takes the form:
\begin{eqnarray}
\beta=\frac{2}{\sigma_s \kappa_1 \lambda_{e-e}}\left\{ G_0 [{\rm Re}g_0^R](\varphi_{I+}^0-2 \tanh \frac{\varepsilon}{2T_e})+ \right.~~~~~~~~ \nonumber \\
\left. G_t [{\rm Re}g_t^R]\varphi_{I-}^0\right\} ~~~~~~~~~~~~\label{beta} \\
\alpha=\frac{2\kappa_1}{\sigma_s (\kappa_1^2-\kappa_2^2) \lambda_s}\left\{ G_0 \left([{\rm Re}g_t^R]-\frac{\kappa_2}{\kappa_1}[{\rm Re}g_0^R]\right)\times \right. ~~~~~~~~~\nonumber \\
\left.(\varphi_{I+}^0-2 \tanh \frac{\varepsilon}{2T_e})+G_t \left([{\rm Re}g_0^R]-\frac{\kappa_2}{\kappa_1}[{\rm Re}g_t^R]\right)\varphi_{I-}^0\right\}
\label{alpha}
~~~~~~
\end{eqnarray}

It is worth to note that for small energies less than the spectral gap these expressions, obtained up to the first order in the I/S interface conductance, are not enough. The exact expressions should be used in order to obtain quantitatively correct answers. It is straightforward to obtain the corresponding formulas from the boundary conditions Eqs.~(\ref{varphi_bc1})-(\ref{varphi_bc2}). We have used the exact expressions in our calculation, but we do not present them here because they are quite cumbersome.     

Further, having at hand the spectral functions $g_{0,t}^R$, obtained from Eq.~(\ref{theta}), the electron temperature $T_e$ from Eq.~(\ref{Te}) and the distribution function $\delta \varphi_\varepsilon = \beta \exp(-\lambda_{e-e}y)$, we can calculate the nonequilibrium spin accumulation $S$ from Eqs.~(\ref{nontherm_spin}) and (\ref{therm_spin}). The results of the calculation are presented in the next section.

\section{results and discussion}

\label{results} 

\begin{figure}[!tbh]
  \begin{minipage}[b]{0.5\linewidth}
     \centerline{\includegraphics[clip=true,width=1.6in]{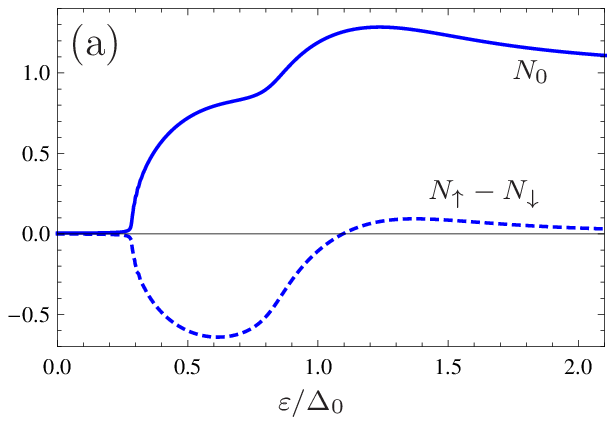}}
     \end{minipage}\hfill
    \begin{minipage}[b]{0.5\linewidth}
   \centerline{\includegraphics[clip=true,width=1.6in]{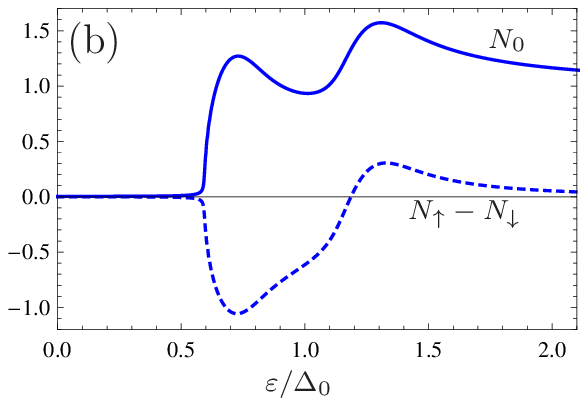}}
  \end{minipage}
\begin{minipage}[b]{0.5\linewidth}
     \centerline{\includegraphics[clip=true,width=1.6in]{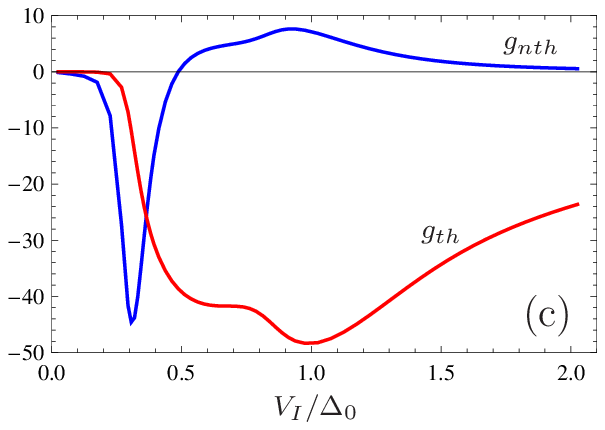}}
     \end{minipage}\hfill
    \begin{minipage}[b]{0.5\linewidth}
   \centerline{\includegraphics[clip=true,width=1.6in]{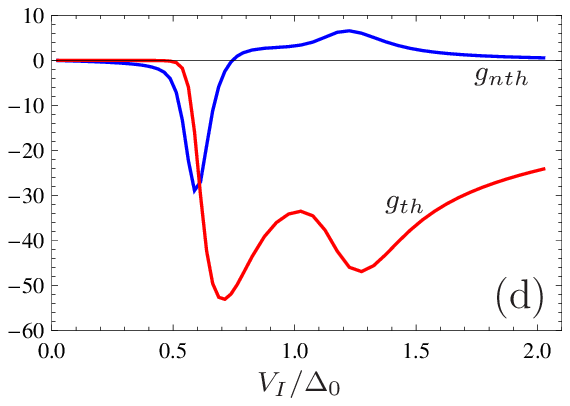}}
  \end{minipage}
   \caption{(a) and (b) spin averaged LDOS $N_0$ (solid line) and the difference between spin-up and spin-down LDOS $(N_\uparrow-N_\downarrow)$ (dashed curve) for (a) $\mu_B H=0.28$, $h_{int}=0$ and (b) $\mu_B H=0.05$, $h_{int} \approx 0.22$; (c) and (d) $g_{nth}$ (blue line) and $g_{th}$ (red line) vs $V_I$ for (c) $\mu_B H=0.28$, $h_{int}=0$ and (d) $\mu_B H=0.05$, $h_{int} \approx 0.22$. All the energies are measured in units of $\Delta_0$.}   
\label{nth_cond}
\end{figure}

We start with a presentation and discussion of the results for nonthermalized contribution to the nonlocal conductance $g_{nth}=dS_{nth}/dV_I$. This quantity vs the injector voltage $V_I$ is presented in Figs.~\ref{nth_cond}(c) and (d). Panel (c) is for the case of no internal exchange field in the film $h_{int}=0$, while panel (d) is for $h_{int}=0.22\Delta_0$. The corresponding spin averaged LDOS $N_0={\rm Re}g_0^R$ and the difference between spin-up and spin-down LDOS $N_\uparrow-N_\downarrow={\rm Re}g_t^R$ are shown in panels (a) and (b) of the same figure, respectively. 

The results represented in the left and right columns manifest approximately the same degree of the Zeeman splitting of LDOS. But, physically, for the left column this Zeeman splitting is entirely provided by the applied field. So te DOS is strongly smeared by the orbital effect of the magnetic field. For the right column the most part of the splitting is due to the internal exchange, so the peaks are more pronounced. It is important that, in spite of this difference, the shapes of the nonlocal conductance $dS_{nth}/dV_I$ exhibit only one peak for the both cases. That is, the Zeeman splitting does not manifest itself in the nonthermalized part of the signal. This is because $g_{nth}(V_I)$ is proportional to $({\rm Re}g_0^R{\rm Re}g_t^R)/(\kappa_1\lambda_{e-e})|_{\varepsilon=V}$, as it can be seen from Eqs.~(\ref{nontherm_spin}) and (\ref{beta}). 

Moreover, the shape of the discussed peak practically does not depend on how particularly the nonthermalized quasiparticles are distributed over the energy levels and mainly determined by the bulk properties of the superconductor. This is strictly valid only if one disregards the dependence $T_e(V_I)$. If it is not the case, this dependence can essentially disturb the peak shape.

\begin{figure}[!tbh]
  \begin{minipage}[b]{\linewidth}
     \centerline{\includegraphics[clip=true,width=2.5in]{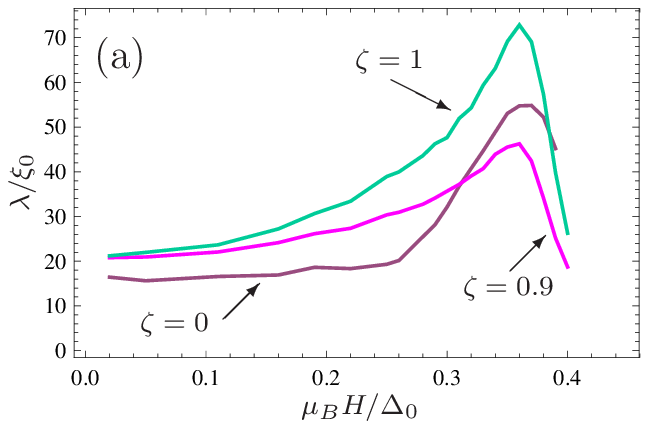}}
     \end{minipage}\hfill
    \begin{minipage}[b]{\linewidth}
   \centerline{\includegraphics[clip=true,width=2.5in]{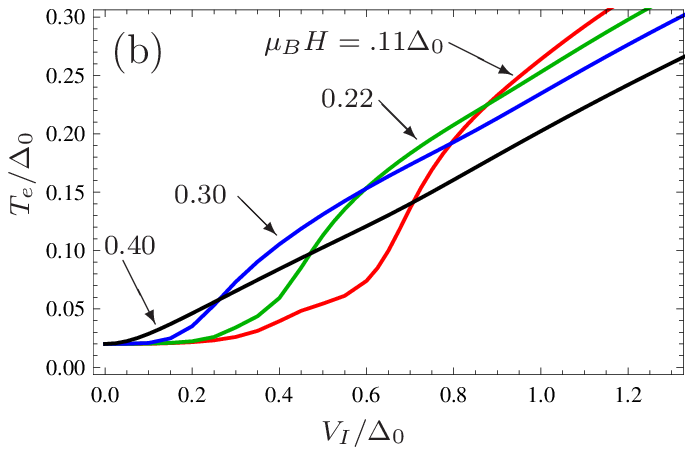}}
  \end{minipage}
   \caption{(a) $\lambda_{e-e}$ vs the applied field ($\zeta=1$) and the relaxation length of the total signal $g_{nth}+g_{th}$ for different values of the phenomenological parameter $\zeta$ (see text); (b) $T_e$ vs $V_I$ for different applied fields. For the both panels $\tau_{so}^{-1}=0.02\Delta_0$, $\tau_{mi}^{-1}=0.01\Delta_0$, $h_{int=0}$, $L_h=80\xi_0$, $G_0/\sigma_s \xi_0=0.002$.}   
\label{e-e_length}
\end{figure}

Now we discuss the relaxation length of the nonthermalized signal. Due to its characteristic one-peak shape the most informative quantity for this purpose is the conductance peak area. One can plot it as a function of $L$ and extract the corresponding relaxation length, as it was done in the experiment \cite{hubler12}. The corresponding relaxation length vs the applied magnetic field is represented in Fig.~\ref{e-e_length}(a) 
for $\zeta =1$. The definition of the parameter $\zeta $ and the physical meaning of the other curves in this figure is explained below in the text. The behavior is very similar to the experimentally observed \cite{hubler12,wolf13}.

According to our theory this typical dependence of the relaxation length on the field is due to the renormalization of the electron-electron relaxation time by superconductivity. Because of the superconducting gap the renormalized $\tau_{e-e}$ is very sensitive to the value of the effective electron temperature $T_e$. This quantity controls the number of the quasiparticles, which appear above the gap and, consequently, available for collisions. The corresponding dependence of $T_e$ on the applied magnetic field is shown in Fig.~\ref{e-e_length}(b). 

In any case $T_e(V_I)$ grows monotonically. Further, it is seen that $T_e$ decreases upon growing the field at any voltage above the spectral gap. The reason for this decrease can be understood in the following way. When the effective electron temperature is smaller than the spectral gap, the number of equilibrium quasiparticles above the gap should be exponentially small. But, from the other hand, this number is controlled by injection. Consequently, the resulting electron temperature is very sensitive to the gap value. The spectral gap shrinks upon the field grows, so  the effective temperature decreases. In addition, when the applied field increases, the characteristic energy, making a larger contribution to the conductance (this energy is $\sim \Delta-h$), goes down. This means that the peak in the nonlocal conductance is formed at lower voltages, where $T_e(V_I)$ is smaller. These are two reasons for the initial growth of $\lambda_{e-e}$ on the applied field. 

The final drop of $\lambda_{e-e}$, seen in Fig.~\ref{e-e_length}(a), has also been observed experimentally \cite{hubler12,wolf13} and can be explained by the fact that the superconducting gap is almost fully suppressed by such large enough magnetic fields, what leads to increase of the number of quasiparticles above the gap. In its turn, this results in sharp increase of the electron-electron relaxation.

The other part of the signal $g_{th}=dS_{th}/dV_I$ is due the electron temperature difference between the film and the detector. This quantity vs the injector voltage $V_I$ is presented in Figs.~\ref{nth_cond}(c) and (d). As it was already discussed in Sec.~\ref{model}, the typical shape of this signal is directly connected to the dependence of electron overheating temperature on the injection voltage. $T_e(V_I)$ has two smeared steps at voltages $V_I \approx \Delta \pm h$, corresponding to the split coherence peaks in the superconducting LDOS. This is because the quasiparticle flow into the superconductor grows sharply at these voltages. Correspondingly, the measured nonlocal differential conductance $dS_{th}/dV_I$ manifests typical two-peak shape, where peaks are located at these voltages. The two peaks can be clearly observed in the signal if the splitting of the coherent peaks in the superconducting DOS is well pronounced.

It can be expected for some setups that the detector is also heated due to injection. In this case the detected spin signal is still determined by Eq.~(\ref{spin}), but $\tanh [{\varepsilon/2T}]$ should be replaced by $\tanh [{\varepsilon/2T_D}]$, where $T_D$ is the detector temperature. Then the thermalized part of the signal can be smaller or even absent. In our work we model this possibility by the phenomenological parameter $\zeta$ as follows: $T_D=T+(T_e-T)\zeta$. That is, $\zeta=0$ for the ideally heat-insulated from the superconductor detector, while $\zeta \to 1$ corresponds to a very good heat contact between the superconductor and the detector. For example, they both can lie on the shared heat-conducting substrate and have a good heat contact to it.     

Because $dS_{th}/dV_I$ has no one distinguished peak, it is not informative to calculate the peak area. So, it is more reasonable to investigate the relaxation length of the signal for a given injector voltage. The relaxation length of this thermalized contribution to the signal is controlled by the length, over which the effective electron temperature relaxes to its equilibrium value. In dependence on the particular sample design it can be determined by the electron-phonon relaxation length or correspond to the distance between the injector and an equilibrium bulk reservoir. In the framework of our model the heat leakage into the phonon subsystem can be neglected and the heat transport is controlled by the temperature gradient. So, it does not depend crucially on any parameters. However, one can imagine another experimental setup, where the distance between the injector and an equilibrium bulk reservoir is extremely large. In this case the resulting relaxation length of $dS_{th}/dV_I$ would be controlled by the electron-phonon relaxation rate. So, it can become very large in this situation, as discussed in Sec.~\ref{phonon}.

\begin{figure}[!tbh]
  \begin{minipage}[b]{\linewidth}
     \centerline{\includegraphics[clip=true,width=2.5in]{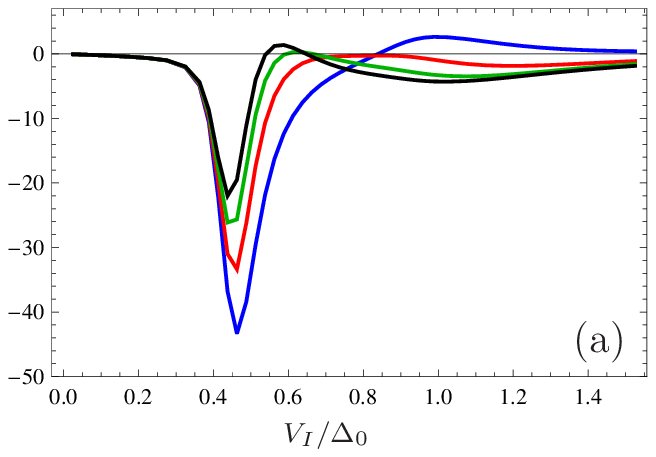}}
    \end{minipage}\hfill
    \begin{minipage}[b]{\linewidth}
   \centerline{\includegraphics[clip=true,width=2.5in]{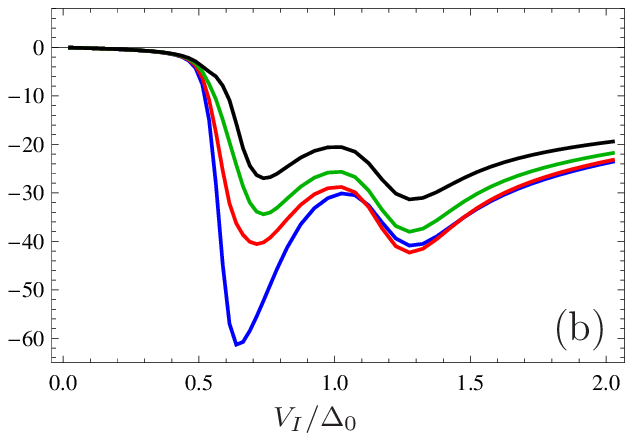}}
  \end{minipage}
\begin{minipage}[b]{\linewidth}
     \centerline{\includegraphics[clip=true,width=2.5in]{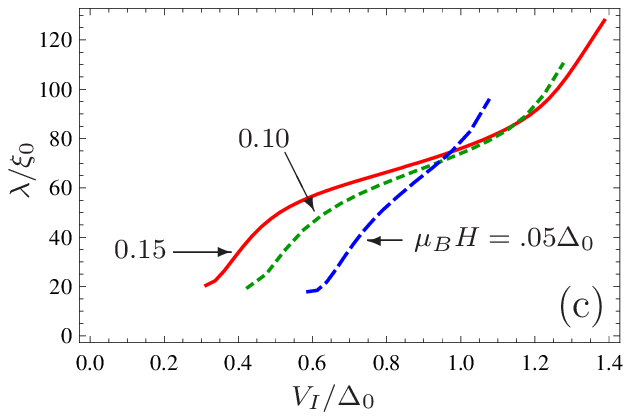}}
     \end{minipage}\hfill
  \caption{(a) and (b) Total conductance vs $V_I$ at different distances from the injector: $L=5,15,25,35 \xi_0$. For (a) $\zeta=0.9$, $\mu_B H=0.22 \Delta_0$, $h_{int}=0$ and the other parameters as in Fig.~\ref{e-e_length}. For (b) $\zeta=0$, $\mu_B H=0.05 \Delta_0$, $h_{int} \approx 0.22 \Delta_0$, $\tau_{so}^{-1}=0.025\Delta_0$, $\tau_{mi}^{-1}=0.01\Delta_0$ and the other parameters as in Fig.~\ref{e-e_length}. (c) Relaxation length of the total conductance vs $V_I$. Different curves correspond to different applied magnetic fields. The other parameters as for panel (b).}   
\label{total_cond}
\end{figure}

In general, the total measured nonlocal conductance has contributions from the both types of signal. Typically it manifests a pronounced peak at $V_I \approx \Delta - h$, provided by as thermalized, so as nonthermalized electrons. The second peak at $V_I \approx \Delta + h$ is only provided by thermalized electrons. The total conductance for $\zeta=0$ (that is, when the contribution of the thermoelectric effect is maximal) is presented in Fig.~\ref{total_cond}(b). The relaxation length of the total conductance $dS/dV_I$ at a given $V_I$ is plotted in Fig.~\ref{total_cond}(c) vs $V_I$. The results are only presented for an interval $V_I$ starting from the first peak position and ending approximately by the second peak position. For lower and higher voltages we are not able to calculate the relaxation length correctly. This is because at lower voltages our $\tau$-approximation for the electron-electron relaxation works not very well, and for higher voltages $T_e$ is too high and the order parameter suppression by this heating should be taken into account in order to obtain the correct relaxation length. 

The typical feature is that the relaxation length grows with voltage increase. The reason is the following. At rather small voltages less than $\Delta$ the signal is dominated by the nonthermalized part, so the relaxation length is governed by the electron-electron relaxation. At larger voltages this part of the signal is already suppressed (see Fig.~\ref{nth_cond}) and the signal is dominated by the contribution from overheating electrons. As it was discussed above, the corresponding length is much larger. So, the growth of the relaxation length upon voltage increase can be viewed as a crossover from electron-electron to electron-phonon dominated relaxation. These finding are in good agreement with the experimental results obtained in Ref.~\onlinecite{wolf14} (where the superconducting film was fabricated on top of a ferromagnetic insulator). In this work very similar two-peak nonlocal conductances and dependencies of the relaxation length on voltage were observed.  

As it was mentioned above, the situations, where $S_{th}$ is very small or even does not seen, can exist. It can be absent if (i) the electron overheating by the injected current is small or (ii) the detector temperature is close to $T_e$. The possible reasons for the situation (ii) are already discussed above. The total conductance for this situation, modeled by $\zeta=0.9$, are represented in Fig.~\ref{total_cond}(a). The relaxation length of the peak  in the total conductance $dS/dV_I$ is plotted in Fig.~\ref{e-e_length}(a) vs the applied field for different values of the phenomenological parameter $\zeta$.  The calculated dependencies of $dS/dV_I$ and the relaxation length vs the magnetic field make us to conclude that the experimental results, obtained in Refs.~\onlinecite{hubler12,wolf13} in the absence of a ferromagnetic insulator under the film, are in good agreement with our theoretical findings for dominating nonthermalized signal, as if the thermoelectric effect is absent or, at least, very small. 

\section{conclusions}

\label{conclusions} 

In conclusion, a theory of long-range spin transport and spin relaxation in Zeeman-split superconducting films at low temperatures is developed. It is suggested that the main mechanism, which determines the relaxation length of a nonequilibrium spin signal in the Zeeman-split superconductors is the spin-independent energy relaxation. 

The spin signal measured by the detector can be divided into two physically different contributions. The first is due to nonthermalized quasiparticle distribution. Its relaxation length is determined by the electron-electron relaxation, renormalized due to superconductivity and grows upon increase of the applied magnetic field.

The second contribution is due to thermalized overheated electron distribution. It is controlled by the difference between the detector temperature and the temperature of the electronic subsystem in the film. This is a thermoelectric effect. In principle, the value of this contribution can be varied experimentally by adjusting the detector temperature. The decay length of this thermoelectric signal is determined by the length on which energy leaves the electronic subsystem and can be very large under special conditions. 

In the framework of our theory the extremely high spin relaxation lengths, experimentally observed in Zeeman-splitted superconductors, and their growth with the magnetic field and with the applied voltage have natural explanations.


\begin{acknowledgments}
The authors are grateful to M. Silaev for many fruitful discussions. The work was supported by Grant of the Russian Scientific Foundation No. 14-12-01290.    
\end{acknowledgments}



\begin{thebibliography}{99}
%
\bibitem{buzdin05}
A.~I.~Buzdin, Rev. Mod. Phys. {\bf 77},  935 (2005).
%
\bibitem{bergeret05}
F.S. Bergeret, A.F. Volkov, and K.B. Efetov, Rev. Mod. Phys. {\bf 77}, 1321 (2005).
%
\bibitem{keizer06}
R.S. Keizer, S.T.B. Goennenwein, T.M. Klapwijk, G. Miao, G. Xiao,
and A. Gupta, Nature (London) {\bf 439}, 825 (2006).
%
\bibitem{robinson10}
J.W.A. Robinson, J.D.S. Witt, and M.G. Blamire, Science {\bf 329}, 59 (2010). 
%
\bibitem{khaire10}
T.S. Khaire, M.A. Khasawneh, W.P Pratt, and N.O. Birge,
Phys. Rev. Lett. {\bf 104}, 137002 (2010).
%
\bibitem{anwar10}
M.S. Anwar, F. Czeschka, M. Hesselberth, M. Porcu, and J. Aarts,
Phys. Rev. B {\bf 82}, 100501(R) (2010).
%
\bibitem{bergeret14}
F. S. Bergeret, I. V. Tokatly, Phys. Rev. B {\bf 89}, 134517 (2014).
%
\bibitem{gomperud15}
I. Gomperud, J. Linder, Phys. Rev. B {\bf 92}, 035416 (2015).
%
\bibitem{jacobsen15_1}
S.H. Jacobsen, J.A. Ouassou, J. Linder, Phys. Rev. B {\bf 92}, 024510 (2015).
%
\bibitem{jacobsen15_2}
S.H. Jacobsen, J.Linder, Phys. Rev. B {\bf 92}, 024501 (2015).
%
\bibitem{giazotto08}
F. Giazotto and F. Taddei, Phys. Rev. B {\bf 77}, 132501 (2008).
%
\bibitem{yang10}
H. Yang, S.-H. Yang, S. Takahashi, S. Maekawa, and S.S.P. Parkin, Nature Mater. {\bf 9}, 586 (2010).
%
\bibitem{poli08}
N. Poli, J.P. Morten, M. Urech, A. Brataas, D.B. Haviland, and V. Korenivski, Phys.Rev. Lett. {\bf 100}, 136601 (2008).
%
\bibitem{hubler12}
F. Hubler, M.J. Wolf, D. Beckmann, and H.v. Lohneysen, Phys. Rev. Lett. {\bf 109}, 207001
(2012).
%
\bibitem{quay13}
C.H.L. Quay, D. Chevallier, C. Bena, M.~Aprili, Nature Phys. {\bf 9}, 84 (2013).
%
\bibitem{wolf13}
M.J. Wolf, F. Hubler, S. Kolenda, H.v. Lohneysen, and D. Beckmann, Phys. Rev. B {\bf 87}, 024517
(2013).
%
\bibitem{wakamura14}
T. Wakamura, N. Hasegawa, K. Ohnishi, Y. Niimi, and YoshiChika Otani, Phys. Rev. Lett. {\bf 112}, 036602 (2014).
%
\bibitem{wolf14}
M. J. Wolf, C. Surgers, G. Fischer, and D. Beckmann, Phys. Rev. B {\bf 90}, 144509 (2014). 
%
\bibitem{silaev14}
M. Silaev, P. Virtanen, T.T. Heikkila, and F.S. Bergeret, Phys. Rev. B {\bf 91}, 024506 (2015).
%
\bibitem{virtanen15}
P. Virtanen, T.T. Heikkila, and F.S. Bergeret, arXiv:1511.00817
%
\bibitem{jedema02}
F.J. Jedema, H.B. Heershe, A.T. Filip, J.J.A. Baselmans, and B.J. van Wees, Nature {\bf 416},  713 (2002).
%
\bibitem{zhao95}
H.L. Zhao and S. Hershfield, Phys. Rev. B {\bf 52},  3632 (1995).
%
\bibitem{morten04}
J.P. Morten, A. Brataas, and W. Belzig, Phys. Rev. B {\bf 70}, 212508 (2004).
%
\bibitem{morten05}
J.P. Morten, A. Brataas, and W. Belzig, Phys. Rev. B {\bf 72}, 014510 (2005).
%
\bibitem{bobkova15}
I.V. Bobkova and A.M. Bobkov, JETP Letters, {\bf 101}, 118 (2015).
%
\bibitem{silaev15}
M. Silaev, P. Virtanen, F.S. Bergeret, T.T. Heikkila, Phys. Rev. Lett. {\bf 114}, 167002 (2015).
%
\bibitem{krishtop15}
T. Krishtop, M. Houzet, J. S. Meyer, Phys. Rev. B {\bf 91}, 121407(R) (2015).
%
\bibitem{machon13}
P. Machon, M. Eschrig, and W. Belzig, Phys. Rev. Lett. {\bf 110}, 047002 (2013).
%
\bibitem{ozaeta14}
A. Ozaeta, P. Virtanen, F.S. Bergeret, and T.T. Heikkila, Phys. Rev. Lett. {\bf 112}, 057001 (2014).
%
\bibitem{kolenda15}
S. Kolenda, M.J. Wolf, and D. Beckmann, arXiv:1509.05568
%
\bibitem{kopnin09}
N.B. Kopnin, Y.M. Galperin, J. Bergli, and V.M. Vinokur, Phys. Rev. B {\bf 80}, 134502 (2009).
%
\bibitem{moor09}
A. Moor, A.F. Volkov, and K.B. Efetov, Phys. Rev. B {\bf 80}, 054516 (2009).
%
\bibitem{nielsen82}
J. Beyer Nielsen, C. J. Pethick, J. Rammer, and H. Smith, J. Low Temp. Phys. {\bf 46}, 565 (1982).
%
\bibitem{usadel}
K.D. Usadel, Phys.Rev.Lett. {\bf 25}, 507 (1970).
%
\bibitem{kuprianov88}
M. Yu. Kuprianov and V. F. Lukichev, Sov. Phys. JETP {\bf 67}, 1163 (1988).
%
\bibitem{bergeret12}
F. S. Bergeret, A. Verso, and A. F. Volkov, Phys. Rev. B {\bf 86}, 214516 (2012).
%
\bibitem{machon14}
P. Machon, M. Eschrig and W. Belzig, Phys. Rev. Lett. {\bf 110}, 047002 (2013);
P. Machon, M. Eschrig and W. Belzig, New Journal of Physics {\bf 16}, 073002 (2014).
%
\bibitem{dimitrova07}
O. V. Dimitrova, V. E. Kravtsov, JETP Letters {\bf 86}, 670 (2008).
%
\bibitem{altshuler}
B.L. Altshuler and A.G. Aronov, in {\it Electron-Electron Interactions in Disordered Systems}, Ed. by A. L. Efros and M. Pollak (Elsevier, New York, 1985).
%
\bibitem{kopnin}
N.B. Kopnin  {\it Theory of Nonequilibrium Superconductivity}, Oxford University Press, 2001.
%
\end{thebibliography}

\end{document}